\documentclass[11.5pt,a4paper,fleqn]{scrartcl}

\usepackage{graphicx} 
\usepackage{amsmath} 
\usepackage{amssymb} 
\usepackage{hyperref} 
\usepackage{authblk} 
\usepackage{natbib} 
\usepackage{psfrag}

\usepackage{framed}
\usepackage{lipsum}
\usepackage{verbatim}
\usepackage{caption}

\usepackage[bottom=1.95in]{geometry}

\title{Bridging the Gap}
\subtitle{Decoding the Intrinsic Nature of Time in Market Data}

\author[1]{James B. Glattfelder}
\author[1]{Anton Golub}

\affil[1]{flov technologies AG, Zug}
\date{\small \today}

\begin{document}

\maketitle

\abstract{
\noindent
Intrinsic time is an example of an event-based conception of time, used to analyze financial time series. Here, for the first time, we reveal the connection between intrinsic time and physical time. In detail, we present an analytic relationship which links the two different time paradigms. Central to this discovery are the emergence of scaling laws. Indeed, a novel empirical scaling law is presented, relating to the variability of what is know as overshoots in the intrinsic time framework. To evaluate the validity of the theoretically derived expressions, three time series are analyzed; in detail, Brownian motion and two tick-by-tick empirical  currency market data sets (one crypto and one fiat). Finally, the time series analyzed in physical time can be decomposed into their liquidity and volatility components, both only visible in intrinsic time, further highlighting the utility of this temporal kinship.
\\ \newline
\textit{Keywords}: scaling laws, physical time, intrinsic time, event-based time, time series analysis, financial data, Brownian motion, ETH/USDT, USD/JPY
}


\section{Introduction}

The flow of time is a central tenet in the subjective perception of reality. Human consciousness is eternally locked in the continuous transition between the past and the future, experienced as the moment of ``now.'' In stark contrast to the experiential familiarity of time, its ontological structure is obscure. From philosophy \citep{mctaggart1908unreality} to physics \citep{glattfelder2019ontological}, the nature of time has been debated for centuries; sometimes, its reality even wholly rejected \citep{connes1994neumann,barbour2001end}. The discovery of time's malleability \citep{einstein1905elektrodynamik} and its resistance to being quantized \citep{dewitt1967quantum} only add to the enigma.

Overall, in economics, the notion of time has only experienced a role of marginal importance. Here, we continue with a tradition of utilizing an alternative conception of time, defined operationally. Within the resulting novel formalism, new insights into the structure of financial time series can be gained. In the following, for the first time, we construct a bridge between the new and old conceptions of time. In other words, a link is uncovered between mathematical properties expressed in physical time and in intrinsic time. Central to this disclosure is the appearance of scaling laws \citep{newman2005power}, old and new. Examples are provided by analyzing empirical tick-by-tick crypto and fiat currency market data, next to Brownian motion.

\section{The Rise of Intrinsic Time}

The idea of modeling financial time series in a new temporal paradigm goes back to
\citep{mandelbrot1967distribution} and has been a reoccurring theme
since \citep{clark1973subordinated,ane2000order,easley2012volume}. In essence, physical time is substituted with an event-based notion of time. This is to say that these novel measures of time are operationally defined using certain intrinsic features of the data being analyzed---as an example, driven by transaction numbers or trading volumes.

In a similar vein, a notable variant of event-based time was introduced in \citep{guillaume1997bird}. The intuition is straightforward: In periods of low
market activity intrinsic time clocks tick slower and, conversely, speed up during
phases of high market activity. The endogenous atoms of intrinsic time are given by what are called directional changes. In a nutshell, these are reversals of
price moves measured from local extremes at different scales (details are given in the next
section). Directional changes are more sophisticated versions of drawups and 
drawdowns found in \citep{pospisil2009formulas}.

Underlying the conventional modeling of financial time series is the assumption
of equidistant time intervals, measured, for instance, in seconds or days. This
rigidity embedded in the application of physical time can be disadvantageous and hide relevant properties. In contrast, the dynamic nature of intrinsic time makes it a much more versatile
tool for analyzing the complexity and heterogeneity of market data.
Within this new paradigm of intrinsic time, novel structures and regularities can be uncovered. For instance, a multitude of scaling laws emerges
\citep{guillaume1997bird,glattfelder2011patterns}, the concept of multi-scale liquidity is introduced \citep{golub2016multi},
 systematic trading strategies can be devised \citep{golub2018alpha}, a variation of the notion of volatility is defined \citep{petrov2019instantaneous}, and an agent-based framework is formulated \citep{petrov2020agent}.
The notion of intrinsic time can be extended to a multi-dimensional
methodology, incorporating more than one financial time series \citep{petrov2019intrinsic}. In \citep{mayerhofer2019three} directional changes and overshoots are analyzed for more general stochastic processes.

One prevailing criticism of the utility of intrinsic time is the methodology's relation to conventional physical time metrics. While many analytic expressions can be derived in the language of intrinsic time, their relationship with concepts known within the context of physical time was missing. We would like to bridge this gap in understanding in the following.

\section{A Brief Review of Theoretical Concepts}

Let the price process $(P_t \colon t \geq 0)$ be governed by Brownian motion $(W_t\colon t \geq 0)$ with volatility $\sigma$. Thus
\begin{equation}
dP_t=\sigma dW_t.
\end{equation}
In practice, one samples the prices over discrete equidistant time intervals of length $\Delta t$. Hence, discrete returns $r(\Delta t)$ are defined as the price differences 
\begin{equation}
\label{eq:ret}
r(\Delta t)=\frac{P(t+\Delta t) - P(t)}{P(t)}.
\end{equation}
In 1990, a scaling law was discovered relating the log returns, sampled over time horizons $\Delta t$, to the time horizon $\Delta t$ \citep{muller1990statistical}. This scaling relationship also holds for the non-logarithmic squared returns, expressed symbolically as
\begin{equation}
\langle r(\Delta t)\rangle_2 \sim \Delta t,
\end{equation}
where $\langle x \rangle _2= \frac{1}{n}\sum_{i=1}^n x_i^2$ is the sample average of the squared values. For Brownian motion it is trivial to validate the scaling law, as
\begin{equation}
\label{eq:expgrw}
\mathbb{E}[r(\Delta t)^2]=\sigma^2 \Delta t.
\end{equation}

The intrinsic time methodology dissects a price curve into directional changes of length $\delta$ with an alternating bimodal direction (up or down). In an up mode, the price either moves above the last local price maximum, updating the maximum, or the difference between the price and the last maximum is evaluated. If this difference exceeds $\delta$, a new directional change is registered and the mode switches to the down mode. The algorithm continues correspondingly. In a shorthand notation, $\delta_{\text{up}}$ and $\delta_{\text{down}}$ denote the up and down directional changes, respectively.

\begin{figure}[t]
    \centering 
    \captionsetup{margin=1.5cm}
    \includegraphics[width=0.875\textwidth]{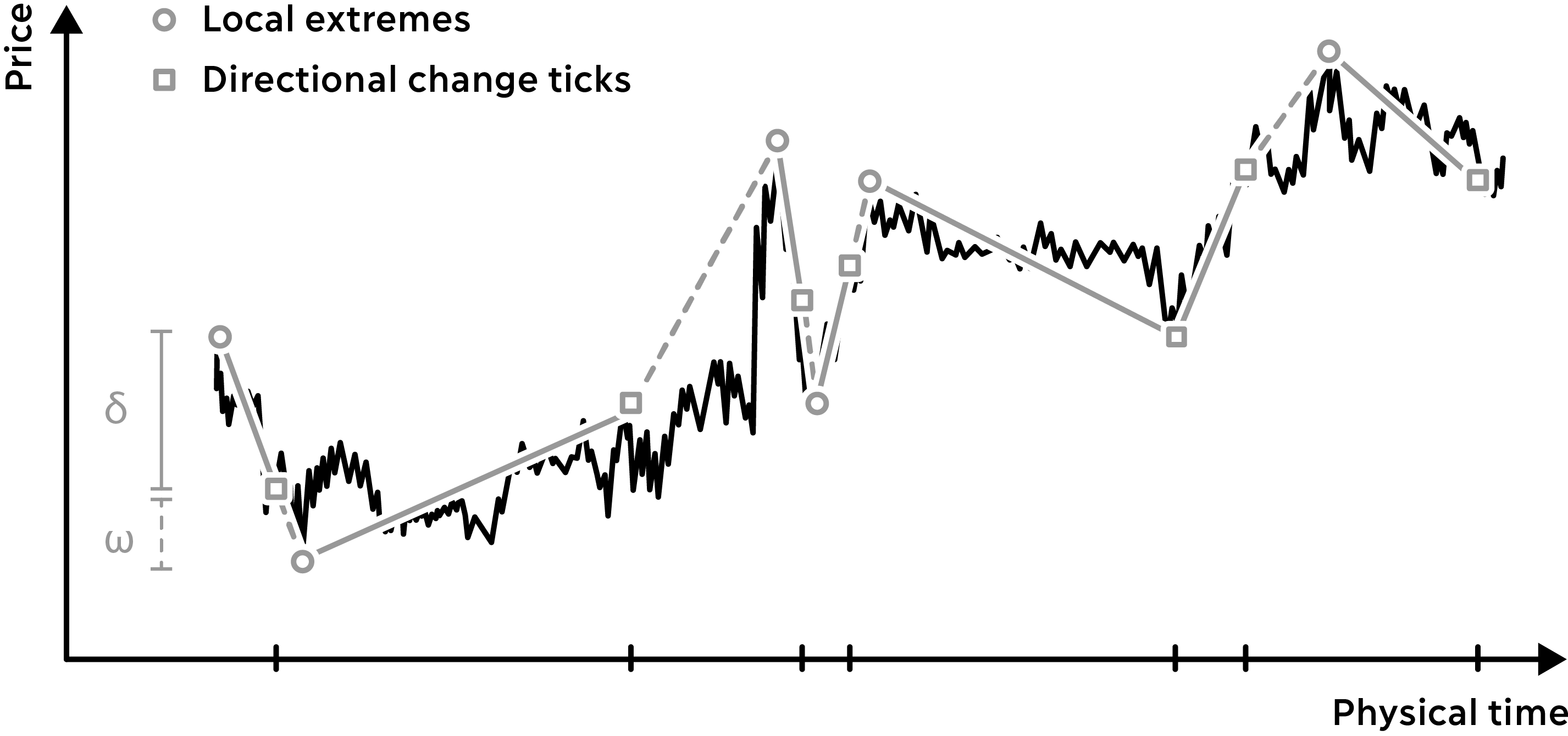}\hspace*{0.5cm}
    \caption{\small Measuring intrinsic time: Starting from the left-hand side, a downward directional change of length $\delta$ is registered and followed by an overshoot
    $\omega$. Each overshoot is defined by the length between the preceding directional change tick and the following local extremum. As a result, a time series can be dissected into directional change and overshoot segments for a given $\delta$. Note that the average overshoot length is given by the scaling law seen in Eq. (\ref{eq:ossl}).
    }
    \label{fig:dcos}
\end{figure}

A directional change can be directly followed by an alternating directional change. However, if the price continues to move in the same direction as the directional change, then what is known as an overshoot $\omega(\delta)$ emerges. As a result, the price curve is dissected into directional change and overshoot segments of the same direction. For instance, for a given $\delta$ the dissected price curve can be: $\delta_{\text{down}}$, $\omega_{1,\text{down}}$,
 $\delta_{\text{up}}$, $\omega_{2,\text{up}}$, $\delta_{\text{down}}$, $\omega_{3,\text{down}}$, \dots While $\delta$ is a fixed threshold, $\omega_{i,\text{\{up,down\}}}$ are of variable length and can be zero. See Fig. \ref{fig:dcos} for further details.

An additional scaling law emerges within the directional change methodology. 
The number of directional changes within a time period $[0,T]$ for a chosen threshold is denoted as $N(\delta, T)$. In \citep{guillaume1997bird} it was established that the number of directional changes scales with the directional change threshold $\delta$. Analytically
\begin{equation}
N(\delta, T)\sim \delta^{-2}.
\end{equation}
For Brownian motion, the expected value of $N(\delta, \sigma, T)$ can be shown to be \citep{petrov2019instantaneous}
\begin{equation}
\label{eq:nodcexp}
\mathbb{E}[N(\delta, \sigma, T)]=\frac{\sigma^2 T}{\delta^2},
\end{equation}
in line with empirical observations. 

The identification of yet another scaling law helped establish the usefulness of
intrinsic time. In detail, 
 the length of the overshoot $\omega(\delta)$ is given by the directional change threshold $\delta$ itself \citep{glattfelder2011patterns}. Hence
\begin{equation}
\label{eq:ossl}
\langle \omega(\delta)\rangle \approx \delta,
\end{equation}
where $\langle x \rangle =\langle x \rangle_1$ is the sample average. 
For Brownian motion, the overshoots are exponentially distributed with the parameter $\delta$
\citep{golub2016multi}
\begin{equation}
\omega(\delta)\sim \operatorname{Exp}(\delta).
\end{equation}
It is thus trivial to obtain the scaling law for this process
\begin{equation}
\label{eq:omegaexp}
\mathbb{E}[\omega(\delta)]=\delta,
\end{equation}
which corresponds to the empirical results. Thus, finally, it should be noted that
\begin{equation}
\label{eq:omegavar}
\operatorname{Var}\left(\omega(\delta)\right)=\delta^2.
\end{equation}
Building on this overshoot scaling law, a liquidity providing trading strategy was formulated \citep{golub2018alpha}. It was also noted, that the overshoot lengths can be related to illiquidity in markets. Specifically, a multi-scale analysis of overshoots can correctly identify and predict liquidity shocks \citep{golub2016multi}.

\section{Building the Bridge}

Assume that we are observing the price evolution over a time horizon $[0,T]$. As seen in Fig. \ref{fig:dcos}, intrinsic time, denoted here by $\tau$, dissects the price curve into directional change and overshoot segments. In effect, this mechanism defines a subordinate process $\sigma W_{\tau}^{[0,T]}$ which records values when an intrinsic time clock ticks.

We chose to express the price evolution through a sequence of upward overshoots followed by downward directional changes and, conversely, downward overshoots followed by upward directional changes. Symbolically
\begin{equation}
\{\omega_{1,\text{up}}, \delta_{\text{down}} \}, \{\omega_{2,\text{down}}, \delta_{\text{up}} \}, \{\omega_{3,\text{up}}, \delta_{\text{down}} \},\dots
\end{equation}
This interpretation can be expressed analytically as
\begin{equation}
\label{eq:osdc}
\sigma W_{\tau}^{[0,T]}=\sum_{i=1}^{N(\delta,\sigma,T)}(-1)^{i}\left(\omega(\delta)-\delta\right).
\end{equation}
In other words, when the subordinate process registers an intrinsic time event, its value is given by the overshoot length offset by the directional change threshold. Hence Eq. (\ref{eq:osdc}) embodies the definition of intrinsic time, as the clock ticks when a reversal of size $\delta$ from the last price overshoot $\omega(\delta)$ is observed. The factor $(-1)^i$ allows for the alternation between upward and downward overshoots. It is, however, irrelevant for the following derivation and can be dropped. Note that $\sigma W_{\tau}^{[0,T]}$ is also a so-called compounded process. 

While the right-hand side of Eq. (\ref{eq:osdc}) expresses the notion of intrinsic time, the process $\sigma W_{\tau}^{[0,T]}$ itself still evolves in physical time as Brownian motion. It thus holds that
 \begin{equation}
 \label{eq:expproc}
 \mathbb{E}[\sigma W_{\tau}^{[0,T]}]=0,
 \end{equation}
 and that
 \begin{equation}
\mathbb{E}[|\sigma{W_{\tau}^{[0,T]}}|^2] = \operatorname{Var}\left(\sigma W_{\tau}^{[0,T]}\right).
\end{equation}

We now utilize an identity found in probability theory. It is based on Wald's equation for the sum of random variables \citep{wald1945some}, generalized for the variance of a compounded processes, also known as the Blackwell-Girshick equation \citep{klenke2006wahrscheinlichkeitstheorie}. The theorem, applied for the second identity, states that
\begin{align}
\label{eq:bg}
\begin{split}
\mathbb{E}[|\sigma & {W_{\tau}^{[0,T]}}|^2] =\operatorname{Var}\left(\sum_{i=1}^{N(\delta,\sigma,T)}(-1)^{i}\left(\omega(\delta)-\delta\right)\right) \\
=& \operatorname{Var}(\omega(\delta)-\delta) \mathbb{E}[N(\delta,\sigma,T)] + 
\mathbb{E}[\omega(\delta)-\delta]^2 \operatorname{Var}(N(\delta,\sigma,T)).
\end{split}
\end{align}
By noting that
\begin{equation}
\operatorname{Var}(\omega(\delta)-\delta) = \operatorname{Var}(\omega(\delta)),
\end{equation}
and, by virtue of Eq. (\ref{eq:omegaexp}),
\begin{equation}
\mathbb{E}[\omega(\delta)-\delta] = 0,
\end{equation}
the expression further reduces to
 \begin{equation}
 \label{eq:master}
     \mathbb{E}[|\sigma{W_{\tau}^{[0,T]}}|^2]=\operatorname{Var}(\omega(\delta)) \mathbb{E}[N(\delta,\sigma,T)].
\end{equation}
In summary, a connection is established between the variance of the subordinate process $\sigma W_{\tau}^{[0,T]}$ and the variability of overshoots and the expected number of directional changes. The only assumption is given in Eq. (\ref{eq:expproc}).

For Brownian motion, the right-hand side can be further simplified by utilizing Eqs. (\ref{eq:nodcexp}) 
and (\ref{eq:omegavar}) to reveal
\begin{equation}
\label{eq:rhs}
\operatorname{Var}(\omega(\delta)) \mathbb{E}[N(\delta,\sigma,T)]=\delta^2 \frac{\sigma^2 T}{\delta^2} =\sigma^2  T.
\end{equation}
It is also found that
\begin{equation}
\label{eq:lhs}
\mathbb{E}[|\sigma{W_{\tau}^{[0,T]}}|^2] = n \sigma^2 \Delta t,
\end{equation}
by setting
\begin{equation}
T = n \Delta t,
\end{equation}
similar to Eq. (\ref{eq:expgrw}). Generally,
the subordinate process over $[0,T]$ behaves like the returns of a price process
$P_t$, sampled over the intervals $\Delta t$ 
\begin{equation}
\sigma{W_{\tau}^{[0,T]}} \equiv \sqrt{n} r (\Delta t).
\end{equation}

In practice, Eq. (\ref{eq:master}) can be validated as follows for any empirical time series. $\mathbb{E}[N]$ is the observed number of directional changes and
$\operatorname{Var}(\omega(\delta)) = \langle \omega(\delta)-\delta\rangle_2$.
Hence, 
\begin{equation}
\label{eq:gap}
\frac{T}{\Delta t}  \langle r(\Delta t)\rangle_2 \approx \langle \omega(\delta)-\delta\rangle_2   N(\delta, T).
\end{equation}
This equality establishes the relationship between the squared returns, sampled equidistantly in physical time, and the building blocks of intrinsic time, namely the variability of overshoots and the number of directional changes. 

It should also be noted that the two intrinsic time expressions on the right-hand 
side of Eq. (\ref{eq:gap}) measure two distinct features of financial time series. The number of directional changes
is a proxy for the volatility of the price process \citep{petrov2019instantaneous}.
Then, the overshoot lengths quantify the liquidity, where longer overshoots correspond
to more illiquid markets \citep{golub2016multi}. In essence, Eq. (\ref{eq:gap}) not only
bridges the gap between physical and intrinsic time, but crucially also 
decomposes time series into their volatility and liquidity components. As a result,
by observing increased squared returns in market data it is now possible to transition
to the intrinsic time framework and identify the source of the change, be it due
to increased volatility, reduced liquidity, or both.

\section{Scaling and Invariance}

\begin{figure}[!t]
    \centering 
    \captionsetup{margin=1.5cm}
    \psfrag{d}{\small $\delta \; [\%]$}
    \psfrag{t}{\small $\Delta t \; [s]$}
    \psfrag{f1}[][tc][1][0]{\small $N(\delta, T)$}
    \psfrag{f2}[][tc][1][0]{\small $\langle r (\Delta t) \rangle_2 \; [\%]$}
    \psfrag{f3}[][tc][1][0]{\small $\langle \omega (\delta) \rangle \; [\%]$}
    \psfrag{f4}[][tc][1][0]{\small $\langle \omega (\delta) - \delta \rangle_2 \; [\%]$}
    \includegraphics[width=0.46\textwidth]{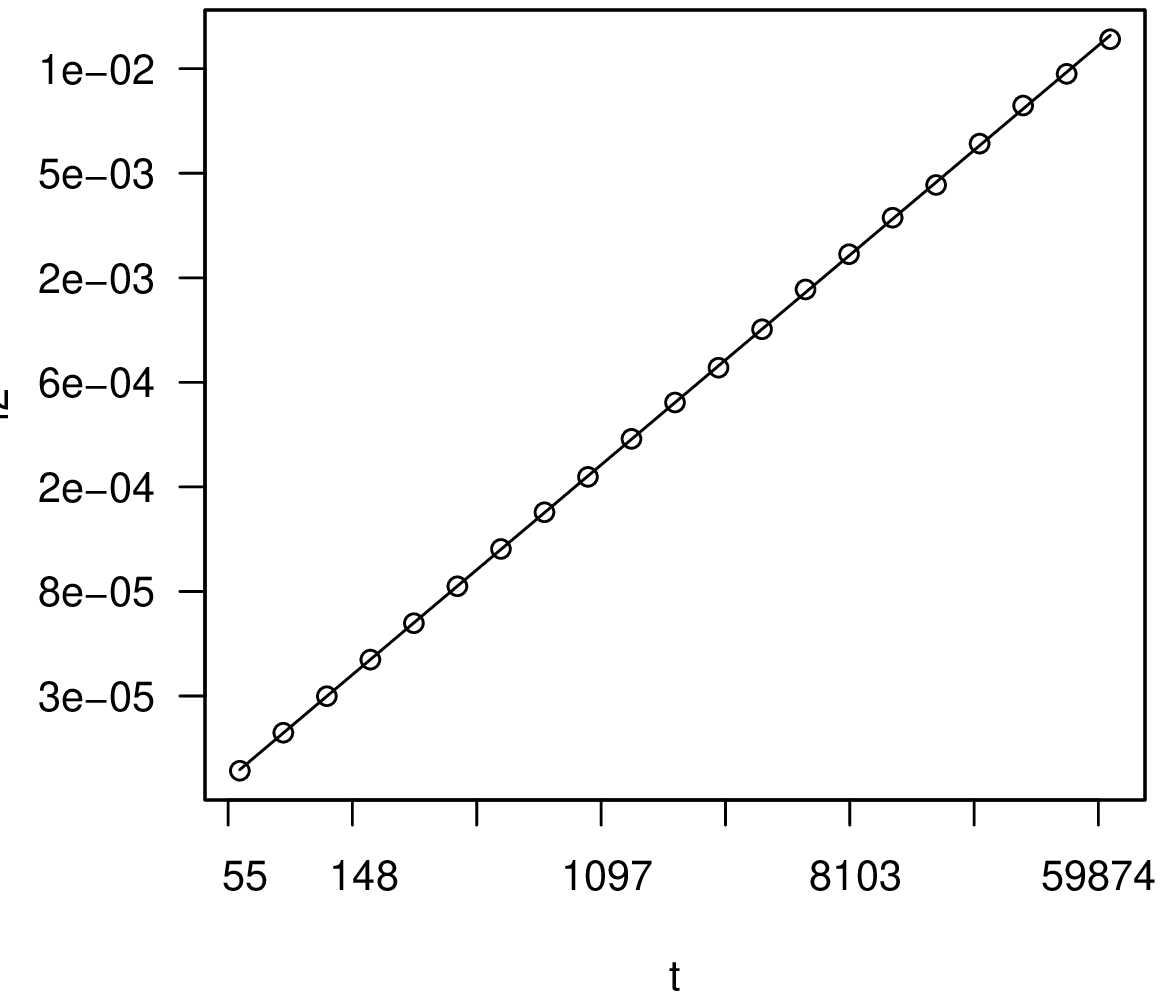} 
    \includegraphics[width=0.46\textwidth]{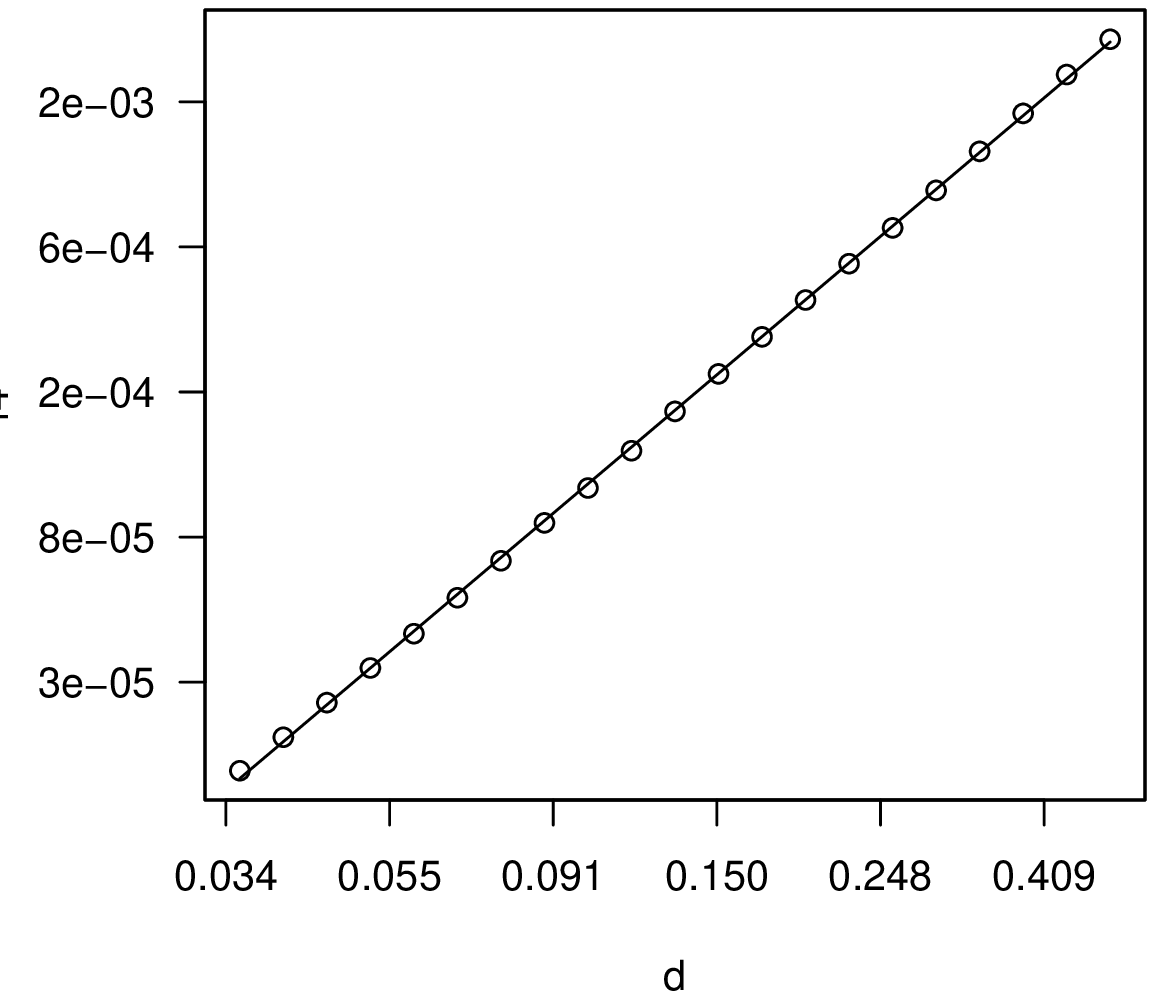}
    \vspace{-0.95cm} \\  
    \includegraphics[width=0.46\textwidth]{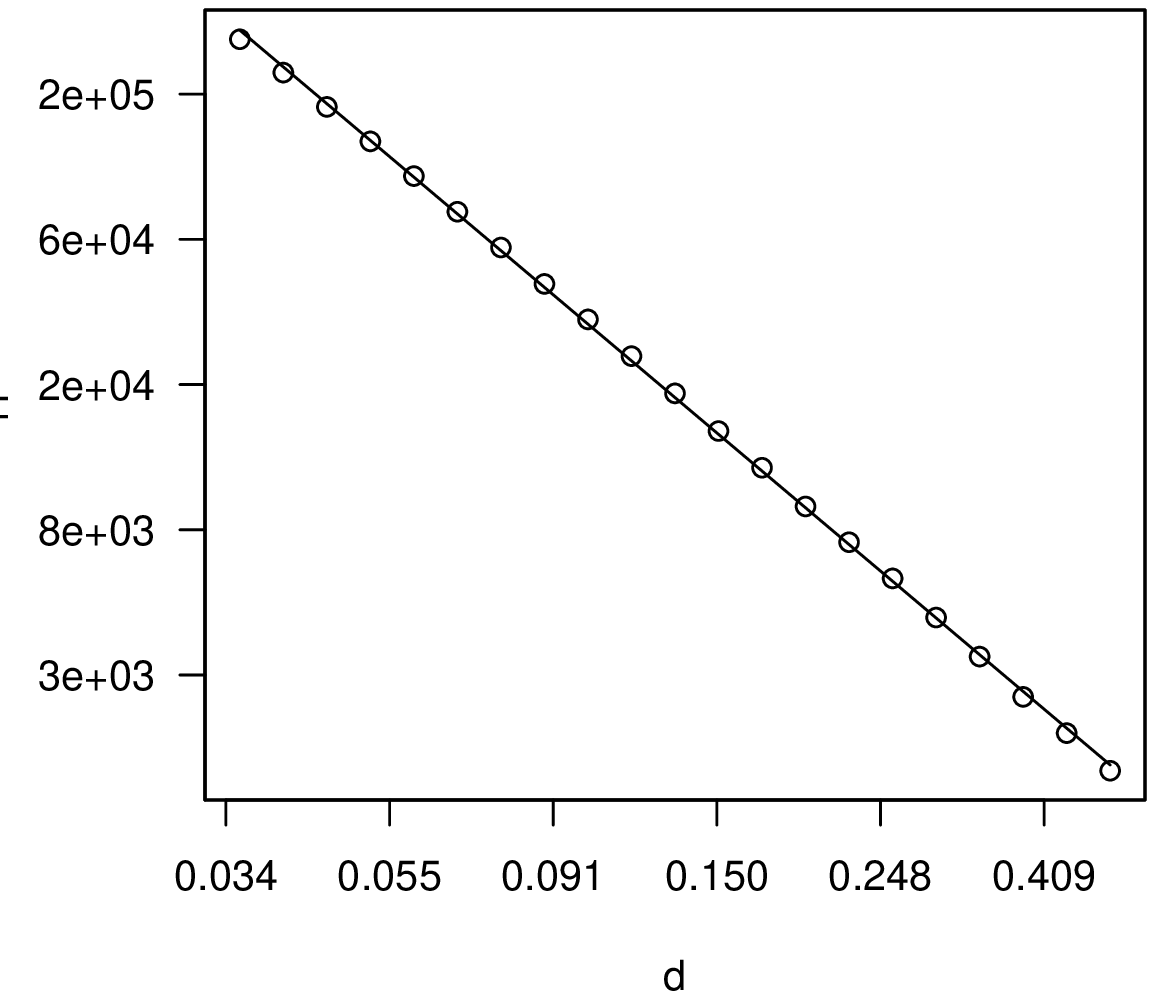}
    \includegraphics[width=0.46\textwidth]{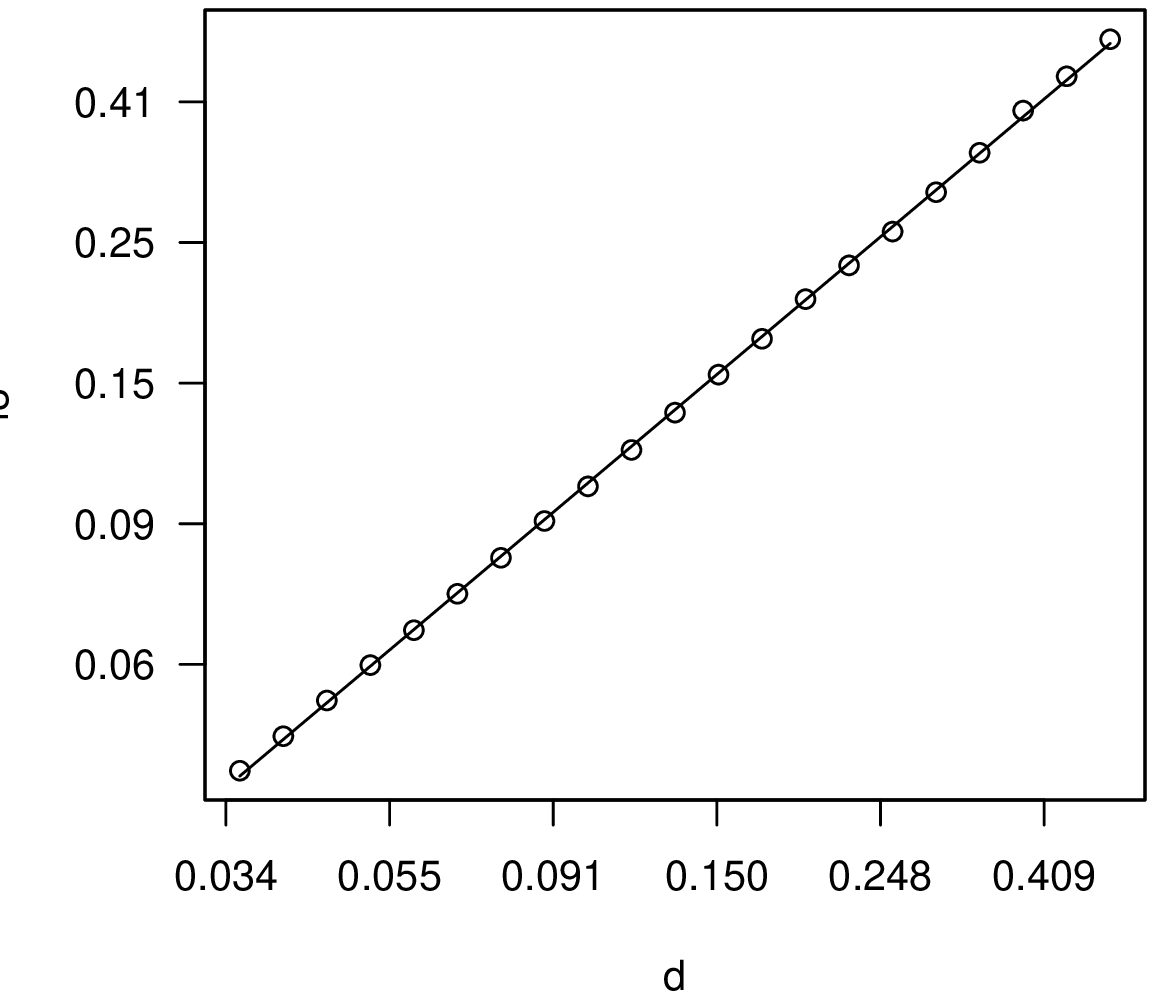}
    \caption{\small Various scaling laws emerging in a Brownian motion time series. See
    Tbl. \ref{tbl:mytabel} for more details.}
    \label{fig:allsls}
\end{figure}

In order to generalize Eq. (\ref{eq:gap}) in its application to other stochastic 
processes and empirical time series, the theoretical scaling behavior of the component 
should be recalled
\begin{align}
&\langle r(\Delta t)\rangle_2 \sim \Delta t, \label{eq:sl1} \\
&N(\delta, T) \sim \frac{T}{\delta^{2}}, \label{eq:sl2}\\
& \langle \omega(\delta)\rangle\sim \delta.
\end{align}
During the analysis performed for this working paper, a novel empirical scaling law was discovered. It relates the variability of overshoots to the directional change threshold
\begin{equation}
    \label{eq:newsl}
    \langle \omega(\delta)-\delta\rangle_2 \sim \delta^2.
\end{equation}
In Figs. \ref{fig:allsls} and \ref{fig:fiatcrypto} empirical examples are plotted. By utilizing Eqs. (\ref{eq:sl1}), (\ref{eq:sl2}), and (\ref{eq:newsl}), both the left-hand side and the right-hand side of Eq. (\ref{eq:gap}) are found to only scale as a function of $T$. Noting that the normalized number of overshoots is given by
\begin{equation}
    \widehat N(\delta) = \frac{N(\delta, T)}{T},
\end{equation}
it holds that
\begin{equation}
\label{eq:inv}
\frac{\langle r(\Delta t)\rangle_2}{\Delta t}   \approx \langle \omega(\delta)-\delta\rangle_2   \widehat N(\delta).
\end{equation}
In effect, the variability due to $\Delta t$ and $\delta$ vanishes from the
expression and an invariant emerges
\begin{equation}
\label{eq:invs}
\mathcal{C}^T := \frac{\langle r(\Delta t)\rangle_2}{\Delta t} \approx
 \langle \omega(\delta)-\delta\rangle_2   \widehat N(\delta) =: \mathcal{C}^{\tau},
 \qquad \forall \; \Delta t \in [t_0, t_T], \delta \in [\delta_0, \delta_\tau],
\end{equation}
where the boundaries of the intervals depend on the length of the time series being analyzed and the directional change thresholds, respectively.

\section{Empirical Analysis}

In order to validate the analytic expressions derived above, one realization of a
Brownian motion time series is analyzed, comprised of 15'631'200 data points spaced
at one second intervals, with $\sigma =  0.00283$. The resulting data set spans close to 181 days. For the computation of the scaling laws, seen in Fig. \ref{fig:allsls}, 21 logarithmically spaced thresholds are chosen:
\begin{itemize}
    \item $\delta$ ranges from 0.035\% to 0.5\%;
    \item $\Delta t$ ranges from 60 seconds to 65'798 seconds, or just over 18 hours.
\end{itemize}
Tbl. \ref{tbl:mytabel} lists the values of the scaling law constants, generically expressed as
\begin{equation}
\label{eq:sl}
    f(x) = \alpha x^E.
\end{equation}

\begin{table}[!t]
\begin{center}
\captionsetup{margin=1.5cm}
\begin{tabular}{lc r r}
    &  & $E$ & $\alpha$    \\ \hline \vspace{-0.4cm}  \\
Squared returns &  $\langle r(\Delta t)\rangle_2 = \alpha  \Delta t^{E}$ & 1.0031 &  $2.4886 \cdot 10^{-9}$   \\
OS variability & $ \langle \omega(\delta)-\delta\rangle_2= \alpha  \delta^{E}$ & 1.9088 &  $6.2213 \cdot 10^{-1}$   \\
Number of normalized DCs & $\widehat{{N}} (\delta) = \alpha \delta^{E}$ & -1.9023 &  $4.3071 \cdot 10^{-9}$   \\
Average overshoot & $\langle \omega(\delta)\rangle = \alpha\delta^{E}$ & 0.9793 &  $9.0140 \cdot 10^{-1}$   \\
\end{tabular}
\caption{Estimation of the empirical Brownian motion scaling law variables, seen in Eq. (\ref{eq:sl}). Fig. \ref{fig:allsls} shows the corresponding charts.}
\label{tbl:mytabel}
\end{center}
\end{table}

\begin{figure}[!b]
    \centering 
    \captionsetup{margin=1.5cm}
    \psfrag{x}{$I$}
     \psfrag{i}{$I$}
    \psfrag{y1}{$\mathcal{C}$}
    \psfrag{f1}{$\mathcal{C}$}
    \psfrag{Li1}{$\mathcal{C}^T$}
    \psfrag{Li2}{$\mathcal{C}^\tau$}
    \includegraphics[width=0.46\textwidth]{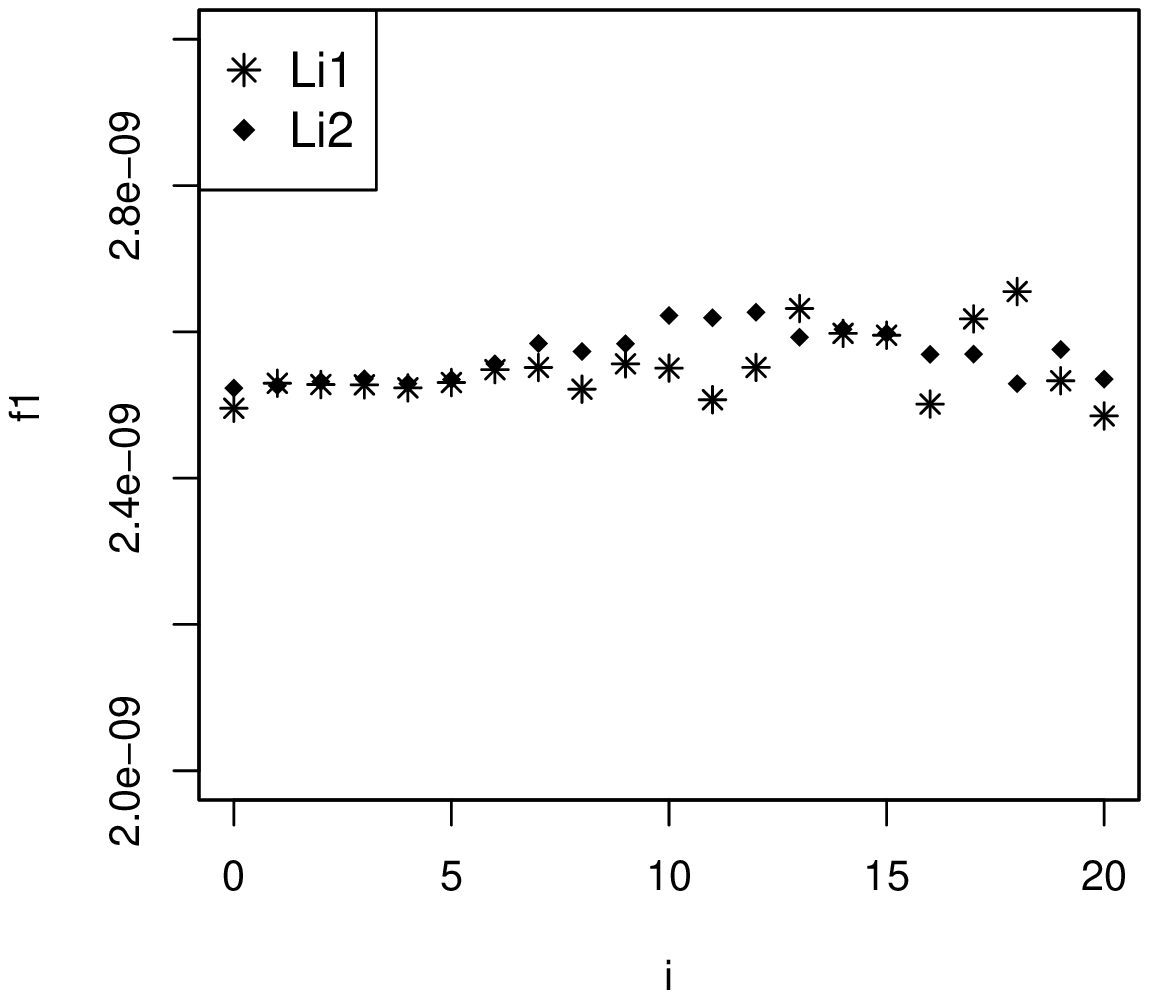}
    \includegraphics[width=0.46\textwidth]{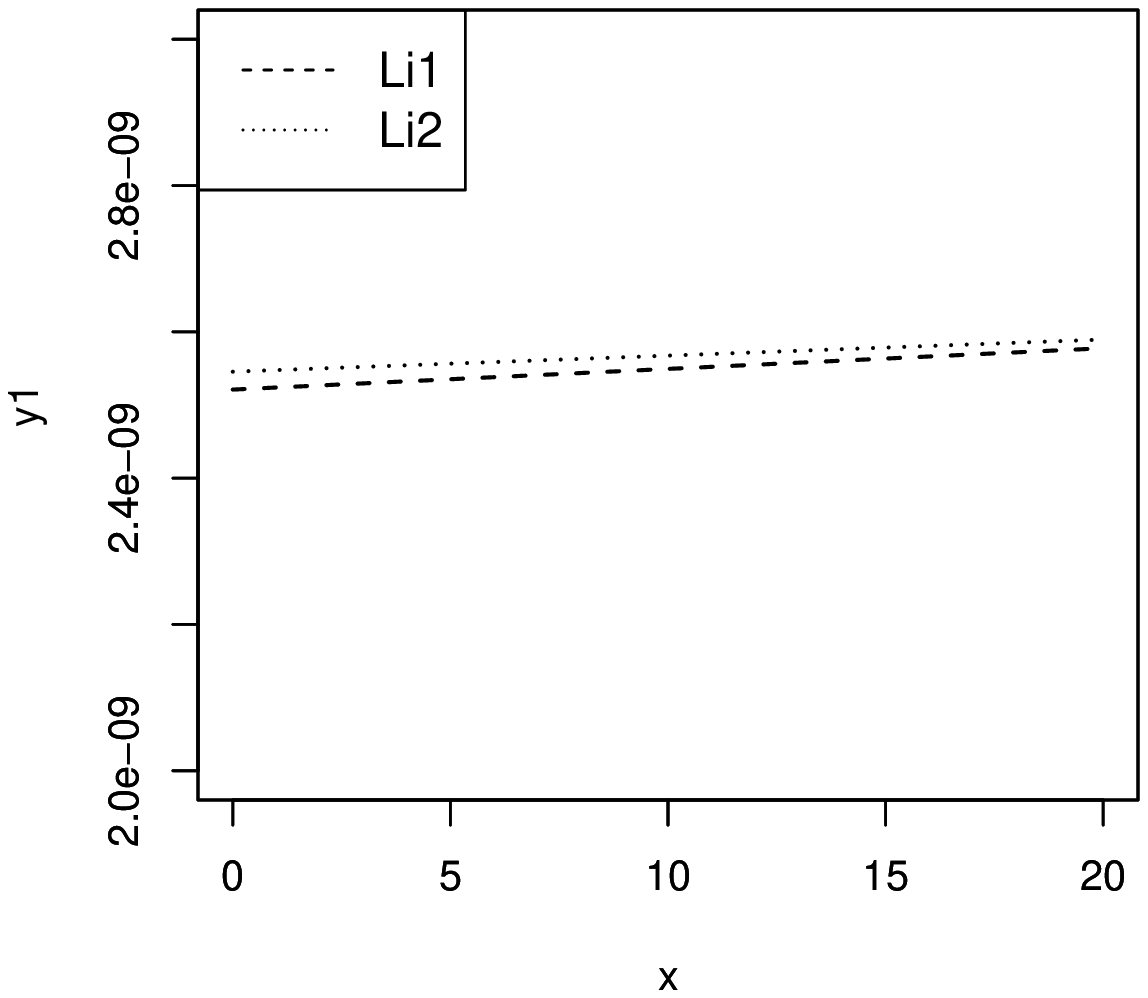}
    \vspace{-0.6cm}
    \caption{\small Evaluation of $\mathcal{C}^{T,\tau}$ for the
   index values $I \in [0, \dots, 20]$ representing the range of thresholds $\Delta t$ and $\delta$, respectively. (\textit{Left}) Scatter plot of all empirical values computed from Eq. (\ref{eq:invs}). (\textit{Right})
   Invariants derived by utilizing the estimated scaling law variables seen in Tbl. \ref{tbl:mytabel}. A deviation from the theoretically expected constant behavior is observed, due to the uncertainty in measurements.}
    \label{fig:C}
\end{figure}

\begin{table}[!t]
\begin{center}
\captionsetup{margin=1.5cm}
\begin{tabular}{l c r r}
 ETH/USDT  & &  $E$ & $\alpha$    \\ \hline \vspace{-0.4cm}  \\
Squared returns  & \phantom{XXXXXXXXXX} & 0.9822 &  $2.5100 \cdot 10^{-8}$   \\
OS variability &  &  1.7249 & $2.4780 \cdot 10^{-1}$   \\
Number of normalized DCs  & & -1.7534 &  $7.3369 \cdot 10^{-8}$   \\
\vspace{-0.4cm}  \\  \vspace{-0.4cm} \\
 USD/JPY   &&  $E$ & $\alpha$    \\ \hline \vspace{-0.4cm}  \\
Squared returns && 0.9813&  $ 8.1806\cdot 10^{-10}$   \\
OS variability && 1.9516 &  $8.4978 \cdot 10^{-1}$   \\
Number of normalized DCs& & -1.8959 &  $ 8.7241\cdot 10^{-10}$   \\
\end{tabular}
\caption{Estimation of the empirical crypto and fiat exchange rate scaling law variables, defined in Eq. (\ref{eq:sl}). See Fig. \ref{fig:fiatcrypto} for the corresponding charts.}
\label{tbl:emp}
\end{center}
\end{table}

In order to compute $\mathcal{C}^T$ from Eq. (\ref{eq:invs}) we choose to evaluate
the squared returns scaling law with $\Delta t = 60$, yielding $\mathcal{C}^T_{60} = 2.5211 \cdot 10^{-9}$. The uncertainty in the empirical realization of Brownian motion results in a numerical deviation from the theoretically expected constant behavior. This is seen, for instance, for the choice of $\Delta t = 65'798$, resulting in $\mathcal{C}^T_{65798} = 2.5776 \cdot 10^{-9}$. For the right-hand side of Eq. (\ref{eq:invs}), the scaling laws evaluated at the boundaries  $\delta \in \{0.035\%, 0.5\%\}$ yield $\mathcal{C}^\tau_\delta \in \{2.5455 \cdot 10^{-9}, 2.5896 \cdot 10^{-9}\}$. The values for $\mathcal{C}^{T,\tau}$ are shown in Fig. \ref{fig:C}. For larger thresholds the uncertainty of $\mathcal{C}^{T} \approx \mathcal{C}^{\tau}$ increases, revealing an empirical imprecision hiding the theoretically expected constant behavior. The average of all empirical realizations is $2.5585 \cdot 10^{-9}$ with a standard deviation of $4.100 \cdot 10^{-11}$.

To further asses the descriptive power of the analytic expressions, we evaluate two tick-by-tick exchange rate time series:
\begin{itemize}
    \item ETH/USDT from 2021-03-01 00:00:00 to 2021-04-15 23:59:59, comprised of 19'324'330 observations and $T=3'974'399$;
    \item USD/JPY from 2013-01-01 00:00:00 to 2013-05-31 23:59:59, comprised of 11'358'048 observations and $T=13'046'399$.
\end{itemize}

\begin{figure}[!h]
    \centering 
    \captionsetup{margin=1.5cm}
    \psfrag{T1}[][c][1][0]{ETH/USDT}
    \psfrag{T2}[][c][1][0]{USD/JPY}
    \psfrag{i}{\small $I$}
     \psfrag{f1}{\small $\mathcal{C}$}
    \psfrag{Li1}{$\mathcal{C}^T$}
    \psfrag{Li2}{$\mathcal{C}^\tau$}
    \includegraphics[width=0.46\textwidth]{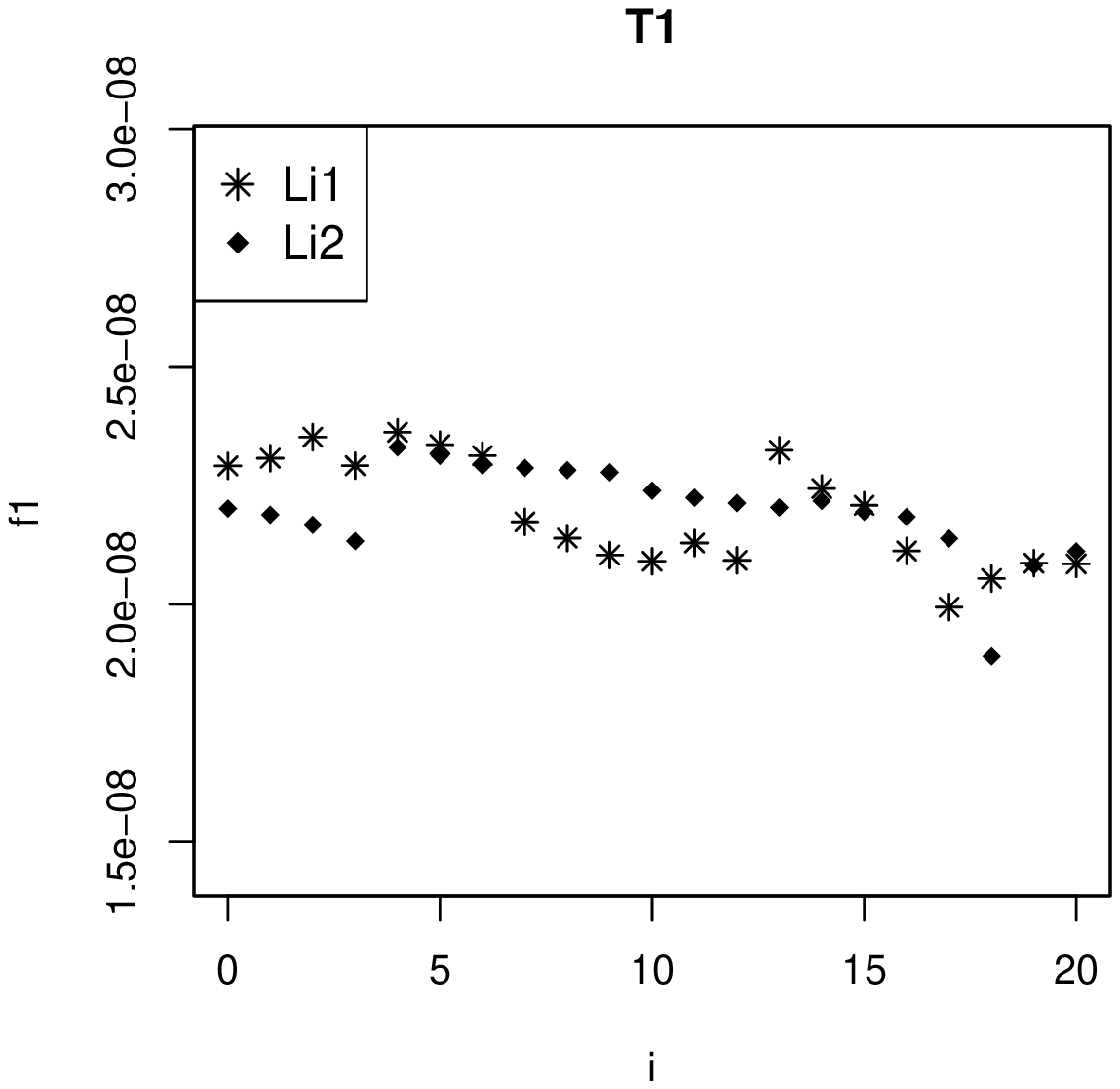} 
    \includegraphics[width=0.46\textwidth]{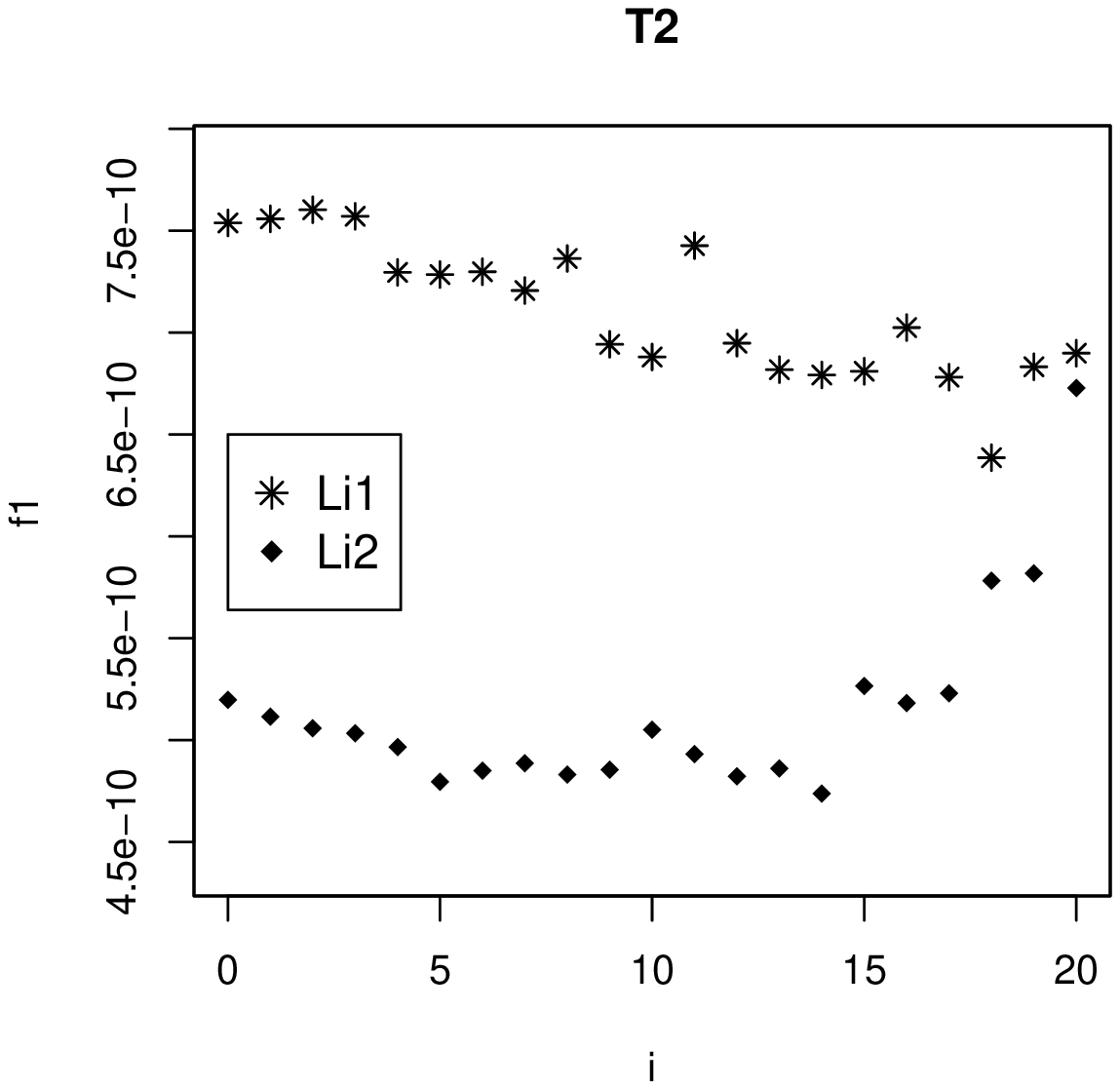}
    \caption{\small Empirical values of $\mathcal{C}^{T,\tau}$ computed from Eq. (\ref{eq:invs}) for the index values $I \in [0, \dots, 20]$. While $\mathcal{C}^T \approx \mathcal{C}^\tau$ for ETH/USDT, for USD/JPY, it appears that $\mathcal{C}^T \neq \mathcal{C}^\tau$. See main text for a discussion.}
    \label{fig:fiatcryptoinv}
\end{figure}

\begin{figure}[!t]
    \centering 
    \captionsetup{margin=1.5cm}
    \psfrag{T2}[][c][1][0]{ETH/USDT}
    \psfrag{T1}[][c][1][0]{USD/JPY}
    \psfrag{d}{\small $\delta \; [\%]$}
    \psfrag{t}{\small $\Delta t \; [s]$}
    \psfrag{f1}[][tc][1][0]{\small $N(\delta, T)$}
    \psfrag{f2}[][tc][1][0]{\small $\langle r (\Delta t) \rangle_2 \; [\%]$}
    \psfrag{f3}[][tc][1][0]{\small $\langle \omega (\delta) \rangle \; [\%]$}
    \psfrag{f4}[][tc][1][0]{\small $\langle \omega (\delta) - \delta \rangle_2 \; [\%]$}
    \includegraphics[width=0.46\textwidth]{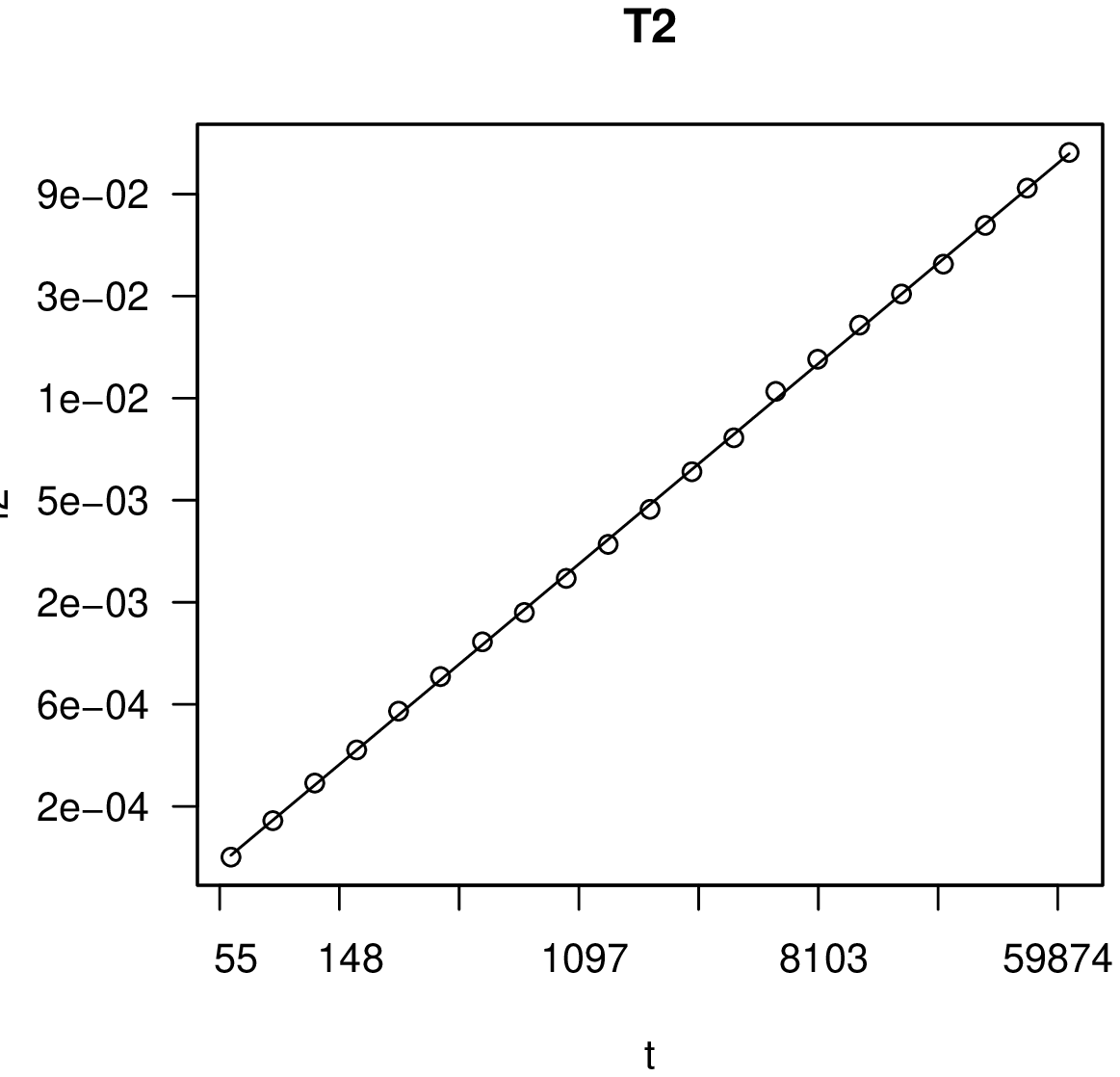} 
    \includegraphics[width=0.46\textwidth]{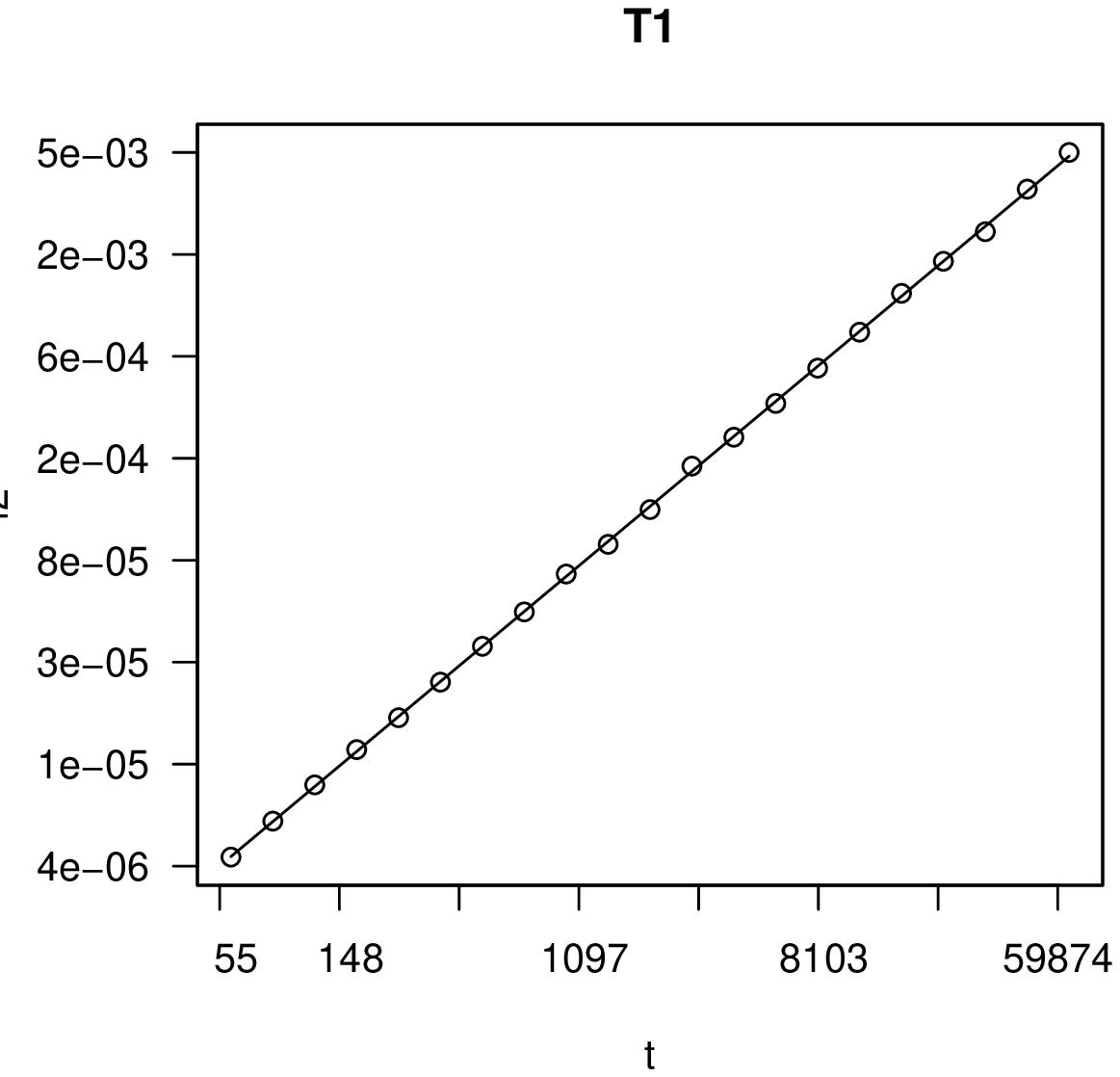}
    \vspace{-0.95cm} \\  
    \includegraphics[width=0.46\textwidth]{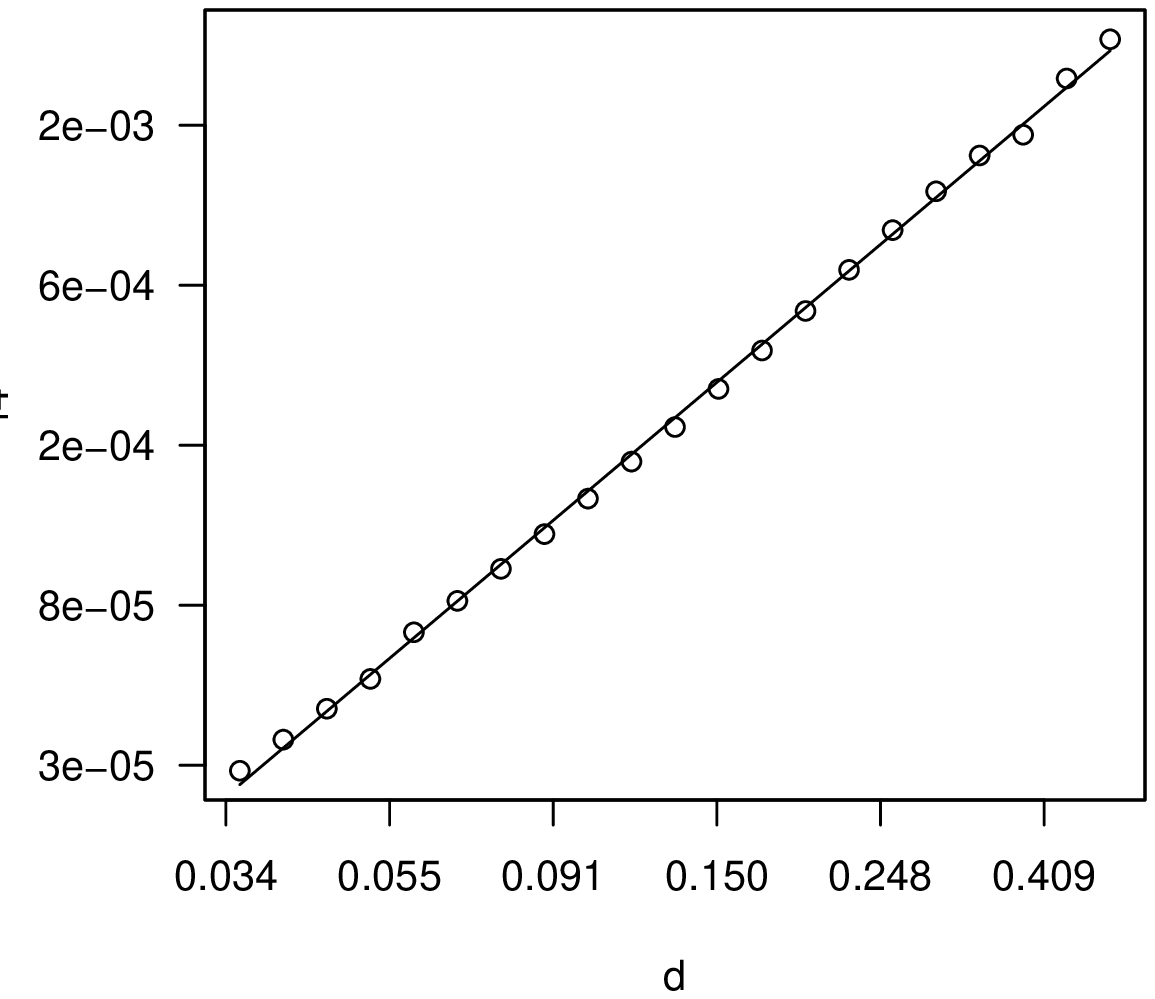}
    \includegraphics[width=0.46\textwidth]{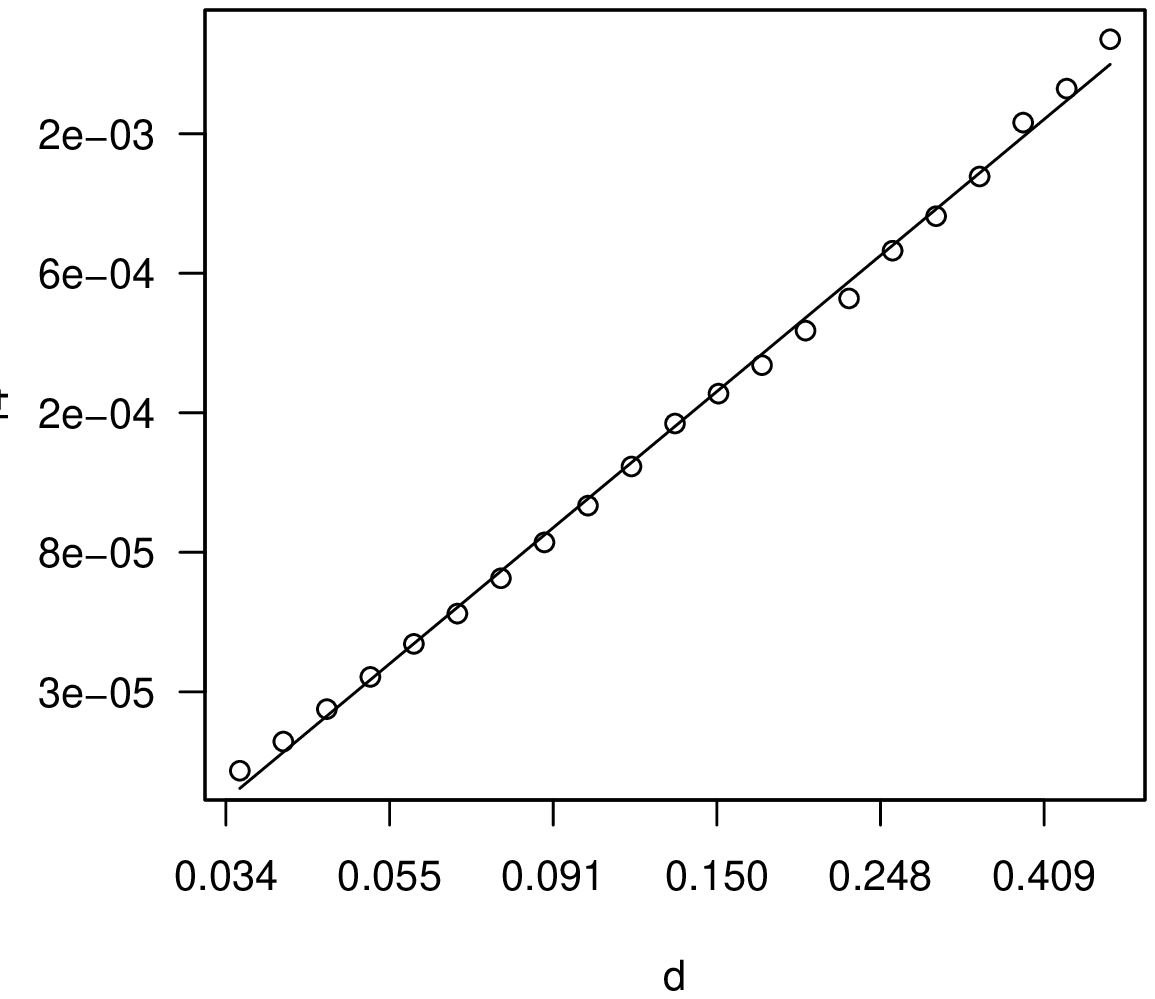}
    \vspace{-0.95cm} \\ 
    \includegraphics[width=0.46\textwidth]{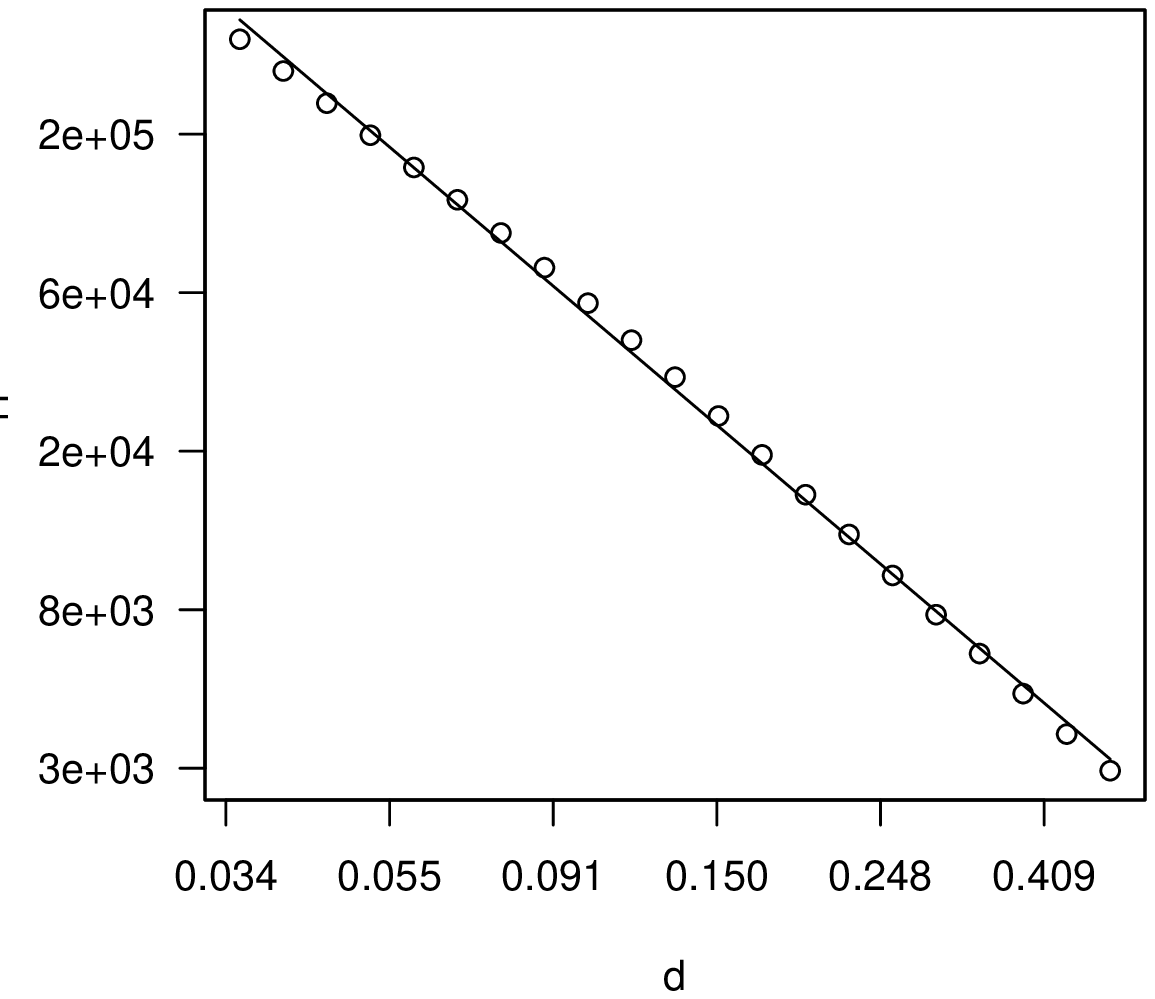}
    \includegraphics[width=0.46\textwidth]{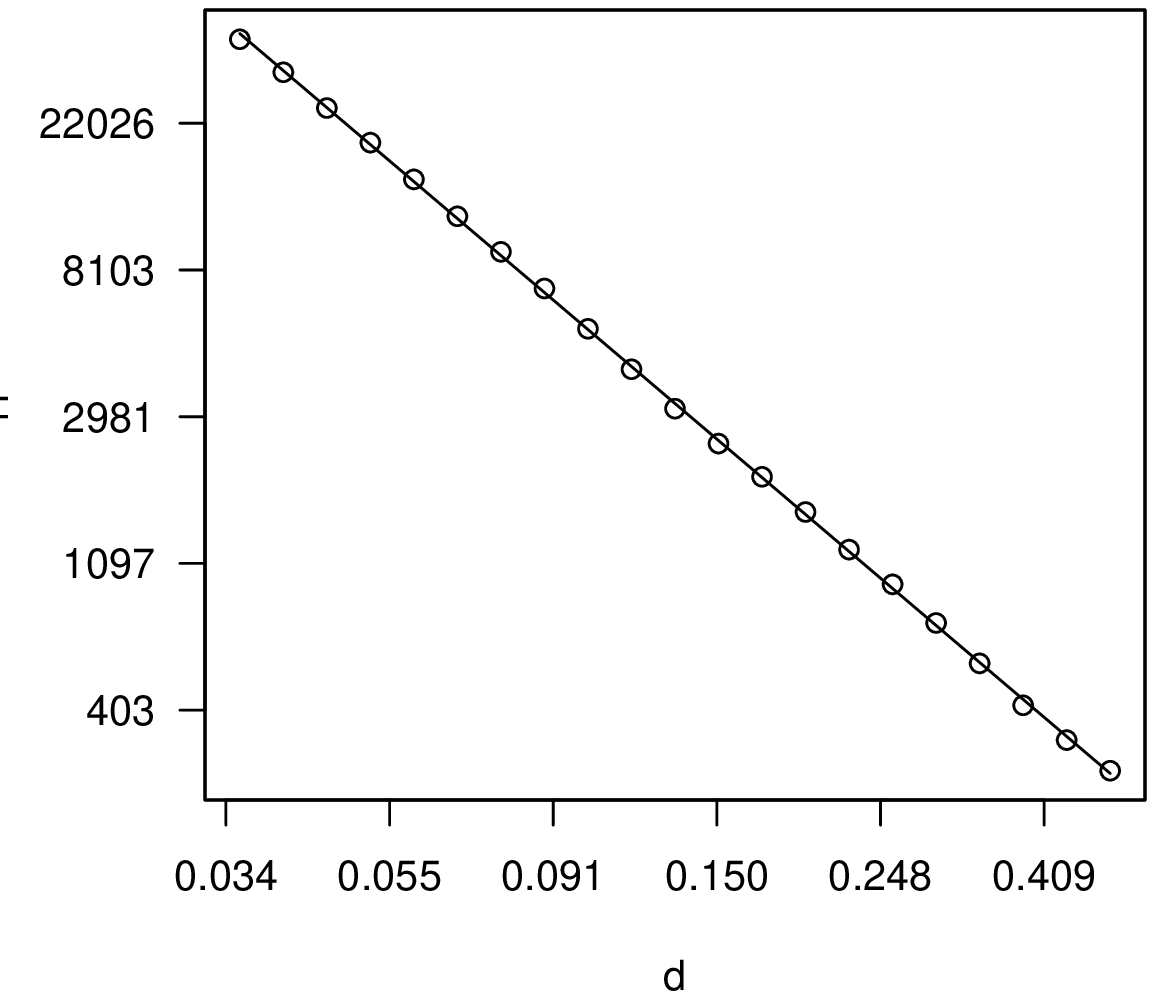}
    \caption{\small Three scaling laws emerging in empirical tick-by-tick market data. (\textit{Left}) A cryptocurrency example using ETH/USDT; (\textit{right}) a fiat currency example with USD/JPY.}
    \label{fig:fiatcrypto}
\end{figure}

In Tbl. \ref{tbl:emp} the estimated scaling law variables are seen and Fig
\ref{fig:fiatcryptoinv} shows the behavior of the invariants $\mathcal{C}^{T, \tau}$. The cryptocurrency invariants for the pair ETH/USDT show greater
variability than what is seen in Fig \ref{fig:C} for Brownian motion,
approximating the constant $\mathcal{C}^{T} \approx \mathcal{C}^{\tau} \approx 2.2094 \cdot 10^{-8}$, with a standard deviation of $ 1.0552 \cdot 10^{-9}$. However, for the fiat 
currency pair USD/JPY Eq. (\ref{eq:invs}) appears to break, and thus Eq. (\ref{eq:gap}) no longer holds. The increased uncertainty for larger thresholds is assumed to be the result of the
higher imprecision due to fewer data points and not due a systematic error. Notably, inspecting the corresponding scaling laws seen in Fig \ref{fig:fiatcrypto} does not reveal a breakdown in the scaling relation. This observation implies the possibility that for
some empirical time series Eq. (\ref{eq:gap}) requires an extension to restore the (approximate) identity. A proposition is to introduce an additional constant scaling 
factor $\lambda$, such that
\begin{equation}
\label{eq:gap2}
\lambda \frac{T}{\Delta t}  \langle r(\Delta t)\rangle_2 \approx \langle \omega(\delta)-\delta\rangle_2   N(\delta, T).
\end{equation}
In the case of USD/JPY one finds that $\lambda  \mathcal{C}^{T} \approx\lambda \cdot 7.107
\cdot 10^{-10} \approx 5.1423 \cdot  10^{-10} \approx   \mathcal{C}^{\tau}$, with $\lambda \approx 0.7235 $.

\section{Conclusion}

In this working paper an analytic relationship between the endogenous and
exogenous conceptions of time is derived. Specifically, measurements in physical time 
are related to measurements performed in intrinsic time. The returns defined in Eq. (\ref{eq:ret}) are a standard metric for analyzing financial time series \citep{dacorogna2001introduction}. Here, the squared returns $\langle r(\Delta t)\rangle_2$ gauge the physical time behavior of the market data. They are linked to measurements in intrinsic time, namely the number of directional changes $N(\delta, T)$ and the variability of overshoots $\langle \omega(\delta)-\delta\rangle_2$. This relationship, expressed in Eq. (\ref{eq:gap}), provides a guideline for connecting the two notions of time. It is found that an increase or decrease in activity observed in physical time can be decomposed into a liquidity and a volatility component, both measured in intrinsic time. Effectively, intrinsic time can uncover novel aspects of time series previously not visible using standard analysis techniques. 

A central element of the analysis is the emergence of scaling laws. Next to three known scaling relations, here, a novel empirical scaling law is uncovered. It relates the variability of the overshoots to the directional change threshold. Now, all three measures used to bridge the gap---the squared returns, the number of directional changes, and the variability of overshoots---can be described by scaling laws. As a result, an invariant is derived for any time series, establishing a taxonomy. In essence, the functional dependency on $\delta$ and $\Delta t$ vanishes.

The theoretical relations can be validated using an idealized time series (Brownian motion) or  empirical tick-by-tick currency market data sets (e.g., ETH/ USDT and USD/JPY). While the uncertainty observed for Brownian motion and the cryptocurrency support the analysis, the fiat currency shows divergent behavior. It is hypothesized, that a novel scaling factor $\lambda$ is required to restore the validity of Eq. (\ref{eq:gap}). In other words, time series can be further classified, depending on their value of $\lambda \neq 1$.

This working paper only glimpses at a possibly rich line of research offered by the notions of physical time, intrinsic time, and their kinship. Further work is required to establish the accuracy of the empirical relationships and the justification for introducing $\lambda$.


\bibliographystyle{apalike} 
\bibliography{ref} 

\end{document}